\documentclass[
reprint,
 amsmath,amssymb,
 aps,
]{revtex4-1}
\usepackage[dvipdfmx]{graphicx}
\usepackage{color}
\usepackage{graphicx}
\usepackage{dcolumn}
\usepackage{bm}
\begin{document}


\title{ Studies of the Electric Dipole Polarizability of $^{129}$Xe using the Relativistic Coupled Cluster and the Normal Coupled Cluster Methods}

\author{Akitada Sakurai$^{1}$}
\author{B. K. Sahoo$^2$}%
\author{B. P. Das$^1$}
\affiliation{
 $^1$Department of Physics, School of Science, Tokyo Institute of Technology, Ookayama, Meguro-ku, Tokyo 152-8550, Japan\\
 $^2$Atomic, Molecular and Optical Physics Division, Physical Research Laboratory, Navrangpura, Ahmedabad 380009, India}

\begin{abstract}
$^{129}$Xe  is an important candidate for the searches of electric dipole moments  due to violations of time-reversal and parity violations in diamagnetic 
atoms. In view of the similarities between the electric dipole moment and the polarizability from the point of view of many-body theory, we have performed
rigorous calculations of the ground state polarizability of $^{129}$Xe using a self consistent relativistic coupled-cluster method and the relativistic normal
coupled-cluster method. The discrepancy between the results from these two methods is two percent, but each of them differs from the accurate measurement 
of the polarizability of the ground state of $^{129}$Xe by one percent. Our results clearly suggest that the two theoretical methods, we have employed in 
this work, can be applied in the future to capture electron correlation effects in the electric dipole moment of $^{129}$Xe to a high degree of accuracy. \end{abstract}

\maketitle

\section{Introduction}
The electric dipole moments (EDMs) of atoms due to violations of time-reversal (T) and parity (P) symmetries are among the leading table-top probes of 
physics beyond the Standard Model of particle interactions \cite{S.M.Barr, M. Pospelov} and they are sensitive to new physics at the TeV scale
\cite{N.Yamanaka}. The EDMs of diamagnetic atoms are primarily sensitive to the nuclear Schiff moment (NSM) and the electron-nucleus tensor-pseudotensor
(T-PT) interaction, which arise from hadronic and semi-leptonic T or CP violation respectively \cite{N.Yamanaka}. A number of experiments are currently 
under way to observe such EDMs \cite{arXiv:1710.02504, arXiv:1803.06821}. The current best EDM limit comes from Hg, which is a diamagnetic atom 
\cite{B. Graner}. Three EDM experiments on another atom of this class, $^{129}$Xe are in progress and new results are expected in the foreseeable 
future \cite{arXiv:1710.02504, arXiv:1803.06821}. These new experimental results for $^{129}$Xe in combination with atomic many-body calculations of the 
ratios of $^{129}$Xe EDM to the NSM and the coupling constant of the T-PT interaction $(C_T)$ separately will yield limits for the NSM and $C_T$.

It is necessary to assess the quality of the atomic many-body calculations of the quantities related to $^{129}$Xe EDM mentioned above. One important 
step in this direction would be to  perform calculations of the ground state electric dipole polarizability of $^{129}$Xe, which has the 
same rank and parity as the  EDM  mentioned above, and therefore both these quantities depend on the same physical effects. The theoretical result obtained 
for $^{129}$Xe polarizability can be compared with its experimental value which has been measured to high accuracy \cite{Hohm U}. These calculations must be 
relativistic in character as $^{129}$Xe is a heavy atom with 54 electrons. Furthermore, it is necessary to use a many-body theory that can capture the 
correlation effects to as high an order as possible in an atom with a large of number of electrons. Taking these two points into consideration, it would 
be appropriate to use the relativistic coupled-cluster (RCC) theory, which is arguably the gold standard for the relativistic theory of atoms and molecules 
\cite{H. S. Nataraj, V. S. Prasannaa}. One important virtue of this theory is that it takes into account correlation effects to all orders in perturbation 
at every level of particle-hole excitation \cite{R. F. Bishop}.  Furthermore it is size-extensive \cite{R. F. Bishop}.

  In the present paper, we have performed rigorous calculations of the electric dipole polarizability of the ground state of $^{129}$Xe using a self-consistent 
RCC method (RCCM) \cite{arXiv:1801.07045} and the relativistic normal coupled-cluster method (RNCCM) \cite{arXiv:1801.07045}. This is the first application 
of the latter method to the calculation of the electric dipole polarizability of the ground state of $^{129}$Xe. The next section gives the salient features 
of these two methods and some key aspects of the calculations. This is followed by a presentation and discussion of our results and finally, we present our 
conclusions.

\section{THEORY AND METHOD OF CALCULATIONS}
 The static polarizability in the uniform dc electric field $\bm{E}$ is defined by
 \begin{equation}
	\langle D \rangle = \alpha {\bm E},	\label{eq:eq_ed}
\end{equation}
where $\langle D \rangle = \langle \Psi_0| D |\Psi_0 \rangle$ is induced electric dipole moment of state $| \Psi_0 \rangle$ of an atom. In the 
first order perturbation, $|\Psi_0 \rangle$  can be expressed as 
\begin{equation}
	|\Psi_0 \rangle = |\Psi_0^{(0)}\rangle +\lambda |\Psi_0^{(1)} \rangle,
	\label{eq:psi_eq}
\end{equation}
where $\lambda$ is perturbed parameter for the Dirac-Coulomb (DC) Hamiltonian given by
\begin{equation}
	H_0^{(DC)} = \sum_{i}^{N_e} [c\bm{\alpha}\cdot \bm{p}_i + m_i c^2 \bm{\beta}+ V_N (r_i) ] +\cfrac{1}{2} \sum_{i,j} \cfrac{1}{r_{ij}},
\end{equation}
and the superscript (0) and (1) represent unperturbed and first-order perturbed wave functions, respectively. In more explicit form, $| \Psi_0^{(1)}\rangle$
can be written as 
\begin{eqnarray}
 	|\Psi_0^{(1)}\rangle &=& \sum_{I} |\Psi_I^{(0)}\rangle \frac{\langle \Psi_I^{(0)} | H_{int}| \Psi_0^{(0)}\rangle}{{E}_0 - { E}_I} \nonumber \\
	&=&  \sum_{I} |\Psi_I^{(0)}\rangle \frac{\langle{ \Psi_I^{(0)} |D| \Psi_0^{(0)}}\rangle}{{E}_0 - {E}_I},
	\label{eq:psi_1_1}
 \end{eqnarray} 
 where $|\Psi_I^{(0)}\rangle$ represents an excited state of $H^{(DC)}_0$, ${E}_0$ and ${E}_I$ are the energies of the ground and excited states, respectively, 
 $\lambda H_{int} = - \bm{D} \cdot \bm{E} $ is a perturbed Hamiltonian, and $\bm{D}$ is the electric-dipole operator. 
 In the above equation, we have used $\lambda H_{int} = - \bm{D} \cdot \bm{E} = DEcos\theta$ and $\lambda = E cos\theta$, 
 where $\theta$ is an angle between ${\bf D}$ and {\bf E}.
 
Using Eqs (\ref{eq:psi_eq}) and (\ref{eq:psi_1_1}),  $\langle \bm{D} \rangle = \langle \Psi_0|\bm{D}|\Psi_0 \rangle$ is written as
\begin{eqnarray}
	\langle \bm{D}\rangle &\simeq& \langle \Psi_0^{(0)}|\bm{D}|\Psi_0^{(0)} \rangle + 2 \lambda \langle \Psi_0^{(0)}|{\bf D}|\Psi_0^{(1)} \rangle \nonumber \\
		&=& 2\sum_{I} \cfrac{ \langle{ \Psi_0^{(0)}| {D} |\Psi_I^{(0)} }\rangle \langle{ \Psi_I^{(0)} | {D} | \Psi_0^{(0)} }\rangle }{{E}_0 - { E}_I}\bm{E},		
		\label{eq:eq_d}
\end{eqnarray}
where the first term does not contribute since the electric dipole operator  $\bm{D}$ is an odd parity operator.
From Eqs (\ref{eq:eq_ed}) and (\ref{eq:eq_d}),  $\alpha$ is given by
\begin{equation}
 	\alpha = -2\sum_{I} \frac{ |\langle \Psi_I^{(0)}|{D}|\Psi_0^{(0)} \rangle|^2}{{E}_0 - { E}_I}.
\end{equation}

\subsection{Unperturbed wave function of  Coupled Cluster Method (CCM)}

\, In the CCM, the unperturbed wave function $|\Psi_0^{(0)}\rangle$ for closed-shell atoms can be expressed as  \cite{I. Shavitt}
 \begin{equation}
 	|\Psi_0^{(0)}\rangle= e^{T^{(0)}} |\Phi_0\rangle
	\label{eq:psi_0}
 \end{equation}
 where $|\Phi_0\rangle$ is the Dirac-Fock (DF) wave function, which is determined using the mean-field approximation and $T^{(0)}$ is the sum of all 
 particle-hole excitation operators. In the coupled-cluster singles and doubles (CCSD) approximation, the excitation operator is $T^{(0)} = T_1^{(0)} + T_2^{(0)}$. In the second quantization notation, these operators can be written as
 \begin{equation}
 	T_1^{(0)} = \sum_{a,i} {t^a_i a_a^{\dagger} a_i}  \mbox{~~ and ~~} T_2^{(0)} = \frac{1}{4}\sum_{a,b,i,j} {t^{ab}_{ij} a_a^{\dagger}a_b^{\dagger}a_j a_i},
 \end{equation}
 where $t^a_i $ and $t^{ab}_{ij}$ are the particle-hole cluster amplitudes, $a_n^{\dagger}$ and $a_n$ are the creation and annihilation operators 
 respectively, and the scripts $ n = a,b$ and $ n = i,j $  represent virtual and occupied orbitals respectively. 
 
To obtain the $T^{(0)}$ amplitudes, we solve the following equations \cite{I. Shavitt}:
 \begin{equation}
  \langle{ \Phi_0^*|({H_N^{DC} e^{T^{(0)}}})_{con}|\Phi_0}\rangle = 0.
  \label{eq:jaco}
 \end{equation}
Here $|\Phi_0^*\rangle$ represents an excited determinantal state with respect to these reference state, $H_N^{DC}$ is the normal ordered Hamiltonian, 
and we use the relation $ e^{-T^{(0)}}H _N^{DC}e^{T^{(0)}} = ({H_N^{DC} e^{T^{(0)}}})_{con}$ with the subscript ``con" representing connected 
terms \cite{I. Shavitt}. In the present work, we have used the Jacobi iterative method to numerically solve Eq. (\ref{eq:jaco}) \cite{https:}.
 
 \subsection{First-order perturbed wave function for the Coupled Cluster Method}
 In the presence of a uniform dc electric field, the atomic Hamiltonian is given by
 \begin{equation}
 	H = H_0^{(DC)} + \lambda H_{int},
 \end{equation}
where the perturbed Hamiltonian is $\lambda H_{int} = - \bm{D} \cdot \bm{E}$ has been define earlier. The first order perturbation equation can be
expressed as 
\begin{eqnarray}
	( H_0^{(DC)} + \lambda H_{int}) (|\Psi_0^{(0)}\rangle +\lambda |\Psi_0^{(1)}\rangle) \nonumber \\
	= (E^{(0)}+\lambda E^{(1)}) (|\Psi_0^{(0)}\rangle +\lambda |\Psi_0^{(1)}\rangle),
\end{eqnarray} 
where $E^{(0)}$ and $E^{(1)}$ are the unperturbed and the first order perturbed energies, respectively. 
Keeping only the first-order terms in $\lambda$ in the above equation, we get
\begin{eqnarray}
	(H_0^{(DC)} - E^{(0)} ) |\Psi_0^{(1)}\rangle
	 &=& - H_{int} |\Psi_0\rangle + E^{(1)} |\Psi_0^{(0)}\rangle \nonumber \\
	 &=&  D  | \Psi_0 \rangle + \langle \Psi_0^{(0)} |H_{int}| \Psi_0^{(0)}\rangle |\Psi_0^{(0)}\rangle \nonumber \\
	 &=&  D | \Psi_0 \rangle,
	 \label{eq:eq_1}
\end{eqnarray}
 where $E^{(1)}$ is zero because $D$ has odd parity. Using the CCM ansatz for closed-shell atoms, we can express the total 
 wave function $|\Psi_0\rangle$, which has a definite parity as
 \begin{equation}
 	|\Psi_0\rangle = e^{T}|\Phi_0\rangle,
	\label{eq:psi_def}
 \end{equation}
where we define
\begin{equation}
	T = T^{(0)} + \lambda T^{(1)},
	\label{eq:t_def}
\end{equation}
where $T^{(1)}$ is the first-order excitation operator due to $H_{int}$. Substituting Eq. (\ref{eq:t_def}) in Eq. (\ref{eq:psi_def}), we get
\begin{equation}
	|\Psi_0\rangle = e^{T^{(0)}+\lambda T^{(1)}} |\Phi_0\rangle = e^{T^{(0)}}(1+\lambda T^{(1)})|\Phi_0\rangle,
\end{equation}
where only terms up to linear in $T^{(1)}$ have been kept. Comparing the above equation with Eq. (\ref{eq:psi_def}), it is clear that the first-order 
wave function can be written as \cite{J. Cizek}
\begin{equation}
	|\Psi_0^{(1)}\rangle = e^{T^{(0)}}T^{(1)} |\Phi_0\rangle.
	\label{eq:psi_1}
\end{equation}
To obtain the $T^{(1)}$ amplitudes, we substitute Eq. (\ref{eq:psi_1}) in Eq. (\ref{eq:eq_1}), and get
\begin{eqnarray}
	\langle \Phi_0^*|e^{-T^{(0)}}H _N^{DC}e^{T^{(0)}} T^{(1)}|\Psi_0 \rangle &=&
	 \langle \Phi_0^*|e^{-T^{(0)}} D e^{T^{(0)}}|\Phi_0\rangle \nonumber \\
	\langle \Phi_0^*|\overline{H_0^{DC}}T^{(1)}|\Psi_0 \rangle  &=& \langle \Phi_0^*|\overline{D}|\Phi_0 \rangle  ,
\end{eqnarray}
where we have used the relation $\bar{A} = e^{-T^{(0)}}Ae^{T^{(0)}}= ({A e^{T^{(0)}}})_{con}$ for the operator $A$ \cite{R. F. Bishop}.

\subsection{CCM expression for polarizability}
\, Using Eqs. (\ref{eq:psi_0}) and (\ref{eq:psi_1}), the expression of the polarizability for the CCM  can be written as \cite{R. F. Bishop} 
\begin{eqnarray}
	 \alpha &=& \frac{\langle \Psi_0|D|\Psi_0 \rangle} {\langle \Psi_0|\Psi_0 \rangle }
	 = 2\frac{\langle \Psi_0^{(0)}|D|\Psi_0^{(1)}\rangle}{\langle \Psi_0^{(0)}|\Psi_0^{(0)} \rangle }\nonumber \\
	 &=& 2\langle \Phi_0|({D^{(0)}} {T^{(1)}})_{con}|\Phi_0 \rangle,
	 \label{eq:alpha_ccm}
\end{eqnarray}
where we define ${D^{(0)}} = e^{{T^{(0)}}^{\dagger}}D e^{T^{(0)}}$. 
In the above equation, we use the connected form of the expectation value for a closed shell atom \cite{I. Shavitt}, which is
non terminating. Therefore in order to calculate the expectation value given in Eq. (\ref{eq:alpha_ccm}), we have used a self-consistent coupled-cluster 
approach in which the combined power of ${T^{(0)}}^{\dagger}$ and $T^{(0)}$ is systematically increased till the result for $\alpha$ converges.

\subsection{Unperturbed wave function of Normal Coupled Cluster Method}

Using the NCCM ansatz,  the unperturbed bra state $\langle \Psi_0^{(0)} |$ can be written as
\begin{equation}
	\langle \widetilde{\Psi}_0^{(0)}| = \langle \Phi_0 |(1+\widetilde{T}^{(0)})e^{-T^{(0)}},
	\label{eq:psi_chi}
\end{equation}
where $T_0$ contains the excitation operators as defined earlier, $\widetilde{T}_0$ is the sum of de-excitation operators and is like $T_0^{\dagger}$. 
Using Eqs (\ref{eq:psi_0}) and (\ref{eq:psi_chi}), we get
\begin{eqnarray}
	\langle \widetilde{\Psi}_0^{(0)}|\Psi_0^{(0)} \rangle 
&=& \langle \Phi_0|(1+\widetilde{T^{(0)}})e^{-T^{(0)}}e^{T^{(0)}}|\Phi_0\rangle \nonumber \\ &=&  \langle \Phi_0|\Phi_0\rangle \nonumber \\
&=& 1.
\end{eqnarray}
Using the above bra state, the expectation value of an one-body operator corresponding to a particular property can be expressed as 
\begin{equation}
	\langle \hat{A} \rangle = \langle \Phi_0 | (1+\widetilde{T}^{(0)})e^{-T^{(0)}} \hat{A} e^{T^{(0)}} | \Phi_0\rangle,
	\label{eq:aki}
\end{equation}
where, A is a general one body operator. The presence of $e^{-T^{(0)}} \hat{A} e^{T^{(0)}}$ ensures that the expression on the right hand side of Eq. (\ref{eq:aki})
terminates. An important attribute of the NCCM is that it satisfies the Hellman-Feynman theorem \cite{R. F. Bishop}. 

To obtain the $\widetilde{T}^{(0)}$ amplitude, we solve the following equation:
\begin{eqnarray}
	\langle \Phi_0|(1+\widetilde{T}^{(0)})[(He^{T^{(0)}})_{con} , C^+_I]|\Phi_0 \rangle = 0,
\end{eqnarray}
here we express as $T^{(0)} = \sum_{I=1}^{Ne} t_I^{(0)} C_I^{+} $, $t_I^{(0)}$ are the amplitudes of the excitations and $C_I^{+}$ represents a string of creation and annihilation operators
corresponding to a given level of particle-hole excitation \cite{R. F. Bishop}.

\subsection{First-order perturbed wave function for NCCM}
\, Similar to $T^{(1)}$, we express the perturbed wave function for the bra state as 
\begin{eqnarray}
	\langle \widetilde{\Psi}_0 | &=&\langle \Phi_0 | (1+\widetilde{T}^{(0)} + 
	\lambda \widetilde{T}^{(1)})e^{-T^{(0)}-\lambda T^{(1)}} 
	\label{eq:nccm_psi}
\end{eqnarray}
In the above expression only terms up to linear in $T^{(1)}$ have been kept,  and $ \widetilde{T}^{(1)}$ is given by $ \widetilde{T}^{(1)} = \sum_{I=1}^{Ne} t_I^{(1)} C_I$.
 
  To obtain  the amplitudes for $\widetilde{T}^{(1)}$,  we solve the following equations:
 \begin{eqnarray}
 \langle \Phi_0|[\widetilde{T}^{(1)}, \overline{H_N} ]|\Phi_0^*\rangle  + \langle \Phi_0|(1+\widetilde{T}^{(0)} )\overline{H_N}|\Phi^*_0 \rangle \nonumber \\ 
  = -  { \langle \Phi_0| [\overline{H_N},(1+\widetilde{T}^{(0)} )T^{(1)}] | \Phi^*_0 \rangle } ,
 \end{eqnarray}
 where $\overline{H_N} = e^{-T^{(0)}}H_N e^{T^{(0)}}$. 
 
 \subsection{NCCM expression for polarizability }
Using Eqs. (\ref{eq:psi_def} ) and (\ref{eq:nccm_psi}), the NCCM expression for polarizability can be written as
\begin{eqnarray}
	\alpha &=& \langle \widetilde{\Psi}^{(0)}_0 |D| \Psi^{(1)}_0 \rangle  + \langle \widetilde{\Psi}^{(1)}_0 |D| \Psi^{(0)}_0 \rangle  \nonumber \\
		&=& \langle \Phi_0|\widetilde{T}^{(1)} \overline{D}|\Phi_0\rangle  + \langle \Phi_0 | (1+\widetilde{T}^{(0)})\overline{D}T^{(1)} |\Phi_0 \rangle 
		\label{eq:polarizability}
\end{eqnarray}
where, we have used relations ${T^{(n)}}^{\dagger}|\Phi_0 \rangle  = 0 $ and $\langle \Phi_0 | {T^{(n)}}=0$, where $n$ is integer. It is clear from the above
expression for polarizability that it terminates  naturally. The NCCM is more versatile than another coupled-cluster approach to properties that was proposed
by Monkhorst \cite{Int. J}. The calculation of atomic polarizabilities by the latter method is less straightforward than that using the NCCM as it would
entail the computation of the double derivative of the energy with respect to the electric field and this would require the knowledge of complicated perturbed 
coupled-cluster amplitudes \cite{A. Shukla}.

\begin{table*}[t]
	\caption{The $\alpha_0$ and $\beta_0$ parameters of the GTOs, which have used in the present calculations.} 
	\begin{tabular}{c||ccccccccc} \hline\hline \\
	Orbital & $s_{1/2}$ & $p_{1/2}$ & $p_{3/2}$ & $d_{3/2}$ & $d_{5/2}$ & $f_{5/2}$ & $f_{7/2}$ & $g_{7/2}$ & $g_{9/2}$ \\\\ \hline \\
	$\alpha_0$ & 0.020422  & 0.042695 & 0.042695 & 0.024227 & 0.024227 & 0.00084 & 0.00084 & 0.0082 & 0.0082  \\\\
	$\beta_0$ & 2.016  &     2.025    &   2.025    &    2.02     &     2.02   &  2.25   &   2.25    &    2.23    &    2.23\\\\ \hline\hline
	\end{tabular}
	\label{tab:alpha_beta}
\end{table*}

\subsection{Error Estimate from triples excitations}\label{sec:Error}

In the present work, the contributions to the polarizability of atomic Xe from three particle-three hole (triple) and higher order excitations have not been included. In order to estimate the size of these neglected effects, we define the following approximate triples RCC amplitudes in a perturbative manner
\begin{equation}
 T_3^{(0),pert}= \frac{1}{3!}\sum_{ijk , abc}  \frac{ ( H_0^{DC} T_2^{(0)})_{ijk}^{abc }}{{\epsilon}_i + {\epsilon}_j+{\epsilon}_k-{\epsilon}_a -{\epsilon}_b -{\epsilon}_c} 
 \label{eq:t30}
\end{equation}
and 
\begin{eqnarray}
 T_3^{(1),pert}= \frac{1}{3!}\sum_{{ijk,abc}}  \frac{ ( H_0^{DC} T_2^{(1)})_{ijk}^{abc}}{\epsilon_i+ \epsilon_j+\epsilon_k-\epsilon_a-\epsilon_b -\epsilon_c} 
 \label{eq:t31}
\end{eqnarray}
with $i,j,k$ and $a,b,c$ subscripts denoting the occupied and unoccupied orbitals, respectively, and $\epsilon$ representing the orbital energies. The contributions of $T_3^{(0),pert}$ will be larger than that of $T_3^{(1),pert}$ as $T_2^{(0)}$ contains physical effects arising in lower order perturbation. In a similar way,
$T_1^{(1)}$ contributions will dominate over those from $T_1^{(0)}$. Based on these considerations, the dominant uncertainty due to the neglected triples excitations are estimated by evaluating 
the expression
\begin{equation}
 \Delta \alpha = 2 \langle \Phi_0 | T_3^{\dagger (0),pert} D T_2^{(0)} T_1^{(1)} | \Phi_0 \rangle.
\end{equation}

\section{Result and Discussions}
In atomic relativistic many-body calculations,  the commonly used basis sets are Gaussian type orbitals (GTOs). In our present work on the polarizability of the xenon atom, we use a two point Fermi nuclear distribution \cite{M. K. Advani}. For a finite size nucleus, the GTOs can represent the natural 
behavior of the relativistic wave functions \cite{Ishikawa Y}.
The radial part of the relativistic wave function using the GTOs are given by 
\begin{equation}
	G_{k}^{L/S} = C_{k}^{L/S} r^{k} e^{-\alpha_k r^2},
\end{equation}
where the index $k = 0,1,2,\cdots$ for $s,p,d,\cdots$ type orbital symmetry, respectively,
 and the index $L(S)$ means the large(small) component of the relativistic wave function. 
Using the kinetic balance condition, we can obtain the radial part of the small component of the wave function from the large component \cite{K. G. Dyall}. 
We have considered 9 relativistic symmetries in the present calculations 
with 40 basis functions for $s_{1/2}$, 39 for both $p_{1/2}$ and $p_{3/2}$, 38 for both $d_{3/2}$ and $d_{5/2}$, 37 for both $f_{5/2}$ and $f_{7/2}$, and 36 for both $g_{7/2}$ and $g_{9/2}$ symmetries.
We have used  even tempered condition for which the exponent $\alpha_i$ can be expressed as $\alpha_i = \alpha_0 \beta_0^{i-1}$ \cite{ALPHA}.
In our calculation, the values of $\alpha_0$ and $\beta_0$ are unique for orbitals  of a given symmetry. The accuracies of  the results for the DF and CCM
calculations depend on these values, (especially $\beta_0$). The DF equations in matrix form are solved for given values of these two parameters and they
are suitably varied so that the energies and the expectation values of $r$, $1/r$ and $1/r^2$ of the occupied orbitals matches with those obtained from the
numerical GRASP2 code \cite{K.G.Dyall et al}. Keeping this value of $\alpha_0$ fixed, the optimal value of $\beta_0$ is obtained by minimizing the DF energy 
as it is derived from the Rayleigh-Ritz variational principle. This leads to
\begin{equation}
	\frac{\partial E_{DF}}{\partial\beta_0} = 0,
\end{equation}
here $E_{DF}$ is total energy at the DF level.
In the present work we have carried out the aforementioned minimization by using the gradient descent method 
\cite{https://www.benfrederickson.com/numerical-optimization/}. The $\alpha_0$ and $\beta_0$ values from this approach are 
listed in Table \ref{tab:alpha_beta}.  

 \begin{table}[t]
 	\caption{Result of static dipole polarizability of $^{129}$Xe in $[ e a_0^3 ]$ .}
	\begin{tabular}{clcll}
	& Method										 &  Our work				& Others							 								\\ \hline
	& DF											 &   26.865 				&	26.87 \cite{arXiv:1710.10946v1}	, 26.918 \cite{singh}	, 26.97 \cite{Latha} 	\\ 
	& CPDF 										 &   26.973 				&	26.98 \cite{arXiv:1710.10946v1}	, 26.987 \cite{singh} 	, 27.7 \cite{Latha}	\\ 
	& LPRCCSD \footnotemark	[1]				 	 &						&	26.432  \cite{Chattopadhyay} \\
	& RCCSD(SC)		  							 &   28.115 				&	28.13 \cite{arXiv:1710.10946v1}									\\ 
	& RNCCSD 									 &   27.508 				&		 														\\ 
	& Experiment		  							 &   27.815(27) \cite{Hohm U}	&		 														\\ \hline
	\end{tabular}
	\footnotetext[1]{Linearized perturbed RCCSD}
	\label{tb:result}
	
	\caption{\,Contributiones of the polarizability of $^{129}$Xe in $[ e a_0^3 ]$  from different terms in RCCSD .}
	\begin{tabular}{c|ccc|cc|c}
	\hline
	Leading Contributions 							&  &   $\alpha$		 \\ \hline
	$(D {T^{(1)}_1} + c.c. )_{con}  $   					&  &   30.416		\\ 
	$ ({T_1^{(0)}}^{\dagger} D T^{(1)}_1 + c.c.)_{con}$  		&  &	-0.376 		\\ 
	$ ({T_1^{(0)}}^{\dagger}D T_2^{(1)} + c.c.)_{con} $  		&  &	 0.115		\\ 
	$( {T^{(0)}_2}^{\dagger} D T^{(1)}_1+ c.c. )_{con}$  		&  &	-3.408 		\\ 
	$({T_2^{(0)}}^{\dagger}D T_2^{(1)} + c.c.)_{con}$		&  &  1.268		\\ \hline
	\end{tabular}
	\label{tb:cont}
\end{table}

\begin{figure}
  \includegraphics[width=8.5cm, height=8.0cm]{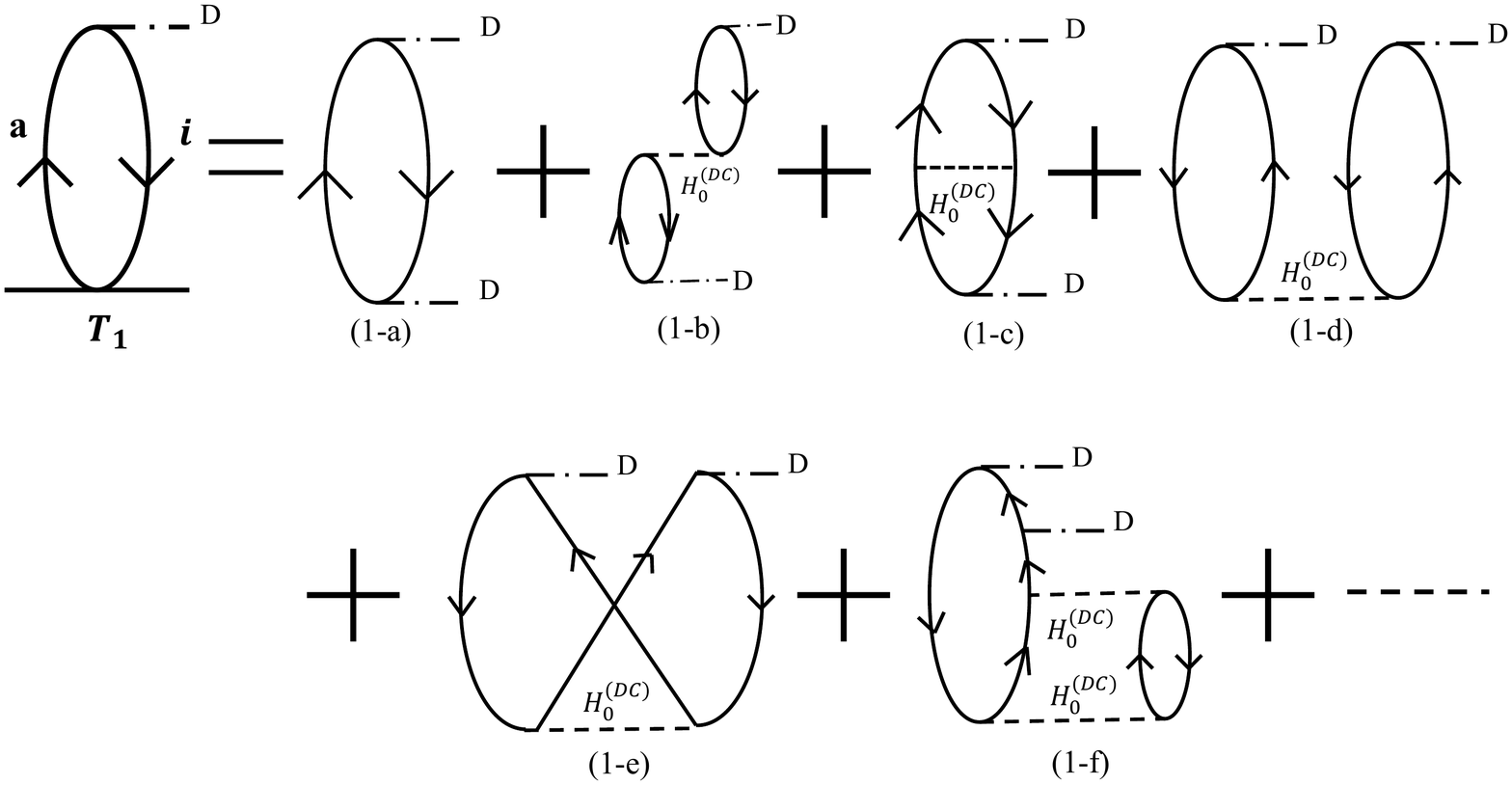}
  \caption{Decomposition of $DT_1^{(1)}$ coupled-cluster diagram into the DF and many-body perturbation theory diagrams.  Here, $D$ and $H^{(DC)}_0$ refer to 
 the dipole and the Dirac-Coulomb (DC) Hamiltonian, which are shown as single dotted and dashed lines, respectively.
  }
   \label{fig:t_1}
\end{figure}

We have performed our polarizability calculations for $^{129}$Xe in the relativistic self-consistent CCSD (RCCSD(SC)) framework and also using the
relativistic NCCSD (RNCCSD) separately. The idea behind the first approach has been stated briefly in the previous section. In
order to make this more transparent, we express Eq. (\ref{eq:alpha_ccm}) as 
\begin{eqnarray}
	\alpha &=& 2\langle \Phi_0|({D^{(0)}} {T^{(1)}})_{con}|\Phi_0 \rangle   \nonumber \\
	&=& 2 \langle \Phi_0 | [ (D+( D T^{(0)}+ c.c) +\cdots) {T^{(1)}}  ]_{con} |\Phi_0\rangle
	\label{eq:alpha_m}
\end{eqnarray}
 in increasing powers of $T^{(0)}$.
 In the self consistent method, $\alpha$  is calculated by increasing successively the combined powers of ${T^{(0)}}^{\dagger}$ and $T^{(0)}$ till  
 self consistency is achieved. The result from the calculations by this method is given in Table \ref{tb:result} . 
 The leading contributions from the terms in Eq. (\ref{eq:alpha_m}) are listed in Table \ref{tb:cont}.
  In Fig. \ref{fig:t_1}, $D T^{(1)}_1$ has been decomposed in terms of the DF, and some lower order many order perturbation theory diagrams. 
It illustrates that a CCM diagram subsumes diagrams corresponding to different physical effects to all orders in perturbation of the residual 
Coulomb interaction.

 In Figs. (1-b) and (1-f) represent typical core polarization and pair correlation effects respectively. From the viewpoint of many-body physics,
the terms in Table \ref{tb:cont} correspond to various kinds of interplay between the core polarization and the pair correlation effects.
The relativistic coupled Hartree-Fock, i.e. the coupled perturbed Dirac-Fock (CPDF) method contains the core polarization effects to all 
orders in the residual Coulomb interaction. Our DF and CPDF results are given in Table \ref{tb:result} and compared with those of other calculations that
were carried out using the same approximations. They are in very good agreement with the results of Refs. \cite{arXiv:1710.10946v1} and \cite{singh}. 
However, our CPDF result differs from that of Ref. \cite{Latha} by about two and a half percent. The reason for this seems to be the different number of 
basis functions and values of the parameters in them that were chosen for the two calculations. All the results for the polarizability calculations given 
in this paper are in atomic units $[ e a_0^3 ]$. In Table \ref{tb:result}, we also give results of different full fledged relativistic coupled-cluster
calculations. Our RCCSD(SC) result is very close to that of another calculation using the same method \cite{arXiv:1710.10946v1}, but with somewhat different 
single particle GTO basis functions. The result of our RNCCSD calculation is also given in Table \ref{tb:result}. The dominant contributions to
$\alpha$ come from $DT^{(1)}_1$ and  $\widetilde{T}^{(1)}_1D$, which arise from $\overline{D}T^{(1)}$ and $\widetilde{T}^{(1)} \overline{D}$, respectively.
These values are 15.208 $(DT^{(1)})$ and 13.180 $(\widetilde{T}^{(1)}D)$ in atomic units (a.u.). The remaining contribution ($-0.88$ a.u.) is due to higher 
order correlation effects that are present in the three terms given in Eq(\ref{eq:polarizability}). The differences in the contributions between the
individual terms of the RCCSD(SC) and their counterparts in the RNCCSD are not negligible. However, the final results for the two methods given in 
Table \ref{tb:result} differ by only two percent. Both of them are in reasonable agreement with an earlier calculation using the RCCSD method which only 
took into account lower order ${T^{(0)}}^{\dagger}$ and $T^{(0)}$ terms for which  the result is 27.744 a.u. \cite{singh}. But they differ from a calculation
based on a linearized perturbed relativistic coupled-cluster singles and doubles (LPRCCSD) approach \cite{Chattopadhyay}  
by about 5 \%. An important reason for this appears to be the non inclusion of correlation effects characterized by the non linear terms in the RCC 
wave function in the latter work.
 
 We identify the three particle-three hole (triples) excitations and the Breit interaction \cite{.P Grant} as the major sources of uncertainties in our 
 polarizability calculations. The error due to the former can be estimated to by calculating the perturbative triple excitations as explained earlier
 in Sec. \ref{sec:Error}. The absolute value of this contribution was found in the present case to be 0.105 a.u.  Given the closeness of the values of $^{129}$Xe
 polarizability at the CPDF and the different coupled-cluster levels (see Table \ref{tb:result}), the Breit interaction for the latter cases can be estimated
 by calculating the contribution of this interaction in the CPDF approximation, and the absolute value obtained for it is 0.051 a.u. The net  uncertainty
 estimated for $^{129}$Xe polarizability calculated by the two variants of RCC theory employed in our present work comes from the
 two above mentioned uncertainties, whose absolute value is 0.156 a.u. for RCCSD(SC). It is reasonable to assume that the uncertainties associated with our 
 RCCSD(SC) and RNCCSD calculations are approximately of the same size; i.e. about 0.6 \% of the total values in the two cases.
 
 \section{Conclusion}
  \, The results of our calculations of the electric dipole polarizability of $^{129}$Xe using the self-consistent relativistic coupled-cluster theory  and
  the relativistic normal coupled-cluster theory have been presented and discussed. They are  within two percent of each other and differ with the measured 
  value by only one percent. The role of correlation effects has been highlighted, and the neglected contributions of these effects and the higher order 
  relativistic effects together are estimated to be about 0.6 \% of the total values of both the relativistic coupled-cluster methods.

  The present work paves  the way for high precision studies of the electric dipole moments of $^{129}$Xe using the two above mentioned relativistic 
  coupled-cluster methods.
 

\end{document}